# Valley polarized edge states beyond inversion symmetry breaking

Dia'aaldin J. Bisharat* & Daniel F. Sievenpiper

Electrical and Computer Engineering Department, University of California San Diego, La Jolla, CA 92093, USA

**Abstract**

Valley-contrast physics has gained considerable attention, particularly for realizing photonic topological insulators (PTIs) that support reflection-free valley-polarized edge modes (VPEMs) in the absence of inter-valley scattering. It is an open question whether similar robust states can exist in the absence of topological valley phase, i.e., nonvanishing Berry curvature at the valleys. We show that a $C_{6v}$-symmetric triangular photonic crystal (PhC) inherently exhibits uniform distribution of spatially varying phase vortices, which support a *local* (limited) version of valley Hall effect (LVHE), where the valley polarization is location defined as opposed to being fixed throughout the bulk. We then demonstrate that defect regions with sublattice asymmetry in otherwise a symmetric PhC lead to wave localization and splitting of photons according to their valley index, thus enabling VPEMs along a line defect waveguide. We fabricate our device on a silicon-on-insulator (SOI) slab and characterize it at near-infrared frequencies showing robust transmission through sharp bends comparable to valley PTIs. Our results present a new perspective to creating valley edge states and outline a new waveguiding mechanism applicable to electromagnetics as well as plasmonics, mechanics and acoustics.

*dbisharat@ucsd.edu

dsievenpiper@ucsd.edu

**Main**

Until recently, routing light around sharp corners in microscopic spaces was generally believed unfeasible except for extremely narrow bands made possible by optimized designs that are unforgiving to fabrication imperfections. This has changed with the advent of PTIs [1-12], which, like their electronic counterparts [13], enable backscattering-immune edge modes along almost arbitrarily shaped interfaces. Among the three basic topological phases –quantum Hall [14], quantum spin-Hall [15, 16], and quantum valley-Hall [17, 18] topological insulators– the latter, which relies on the valley degree of freedom (DOF) [19] in crystals is arguably the easiest to implement in bosonic systems [20-30] and is of the most interest to the development of on-chip optical devices, motivating this work. Generally, it is assumed that a valley-contrast response requires a transition to a topological valley phase, and so a reduction of the lattice symmetry to $C_{3v}$ symmetry. In addition, VPEMs have only been observed at the interface between valley PTIs with opposite valley Chern numbers. Here we assess the degree to which these caveats are practically necessary, demonstrating valley-contrast response and robust VPEMs in the absence of these criteria.

**Local valley Hall effect**

We consider a two-dimensional (2D) dielectric PhC slab with a triangular lattice of circular holes as shown in Fig. 1a. This PhC is widely used for waveguide applications due to its inherent bandgap for TE modes. The dashed line rhombus illustrates the Wigner–Seitz unit cell of the lattice with vertices located at the holes' centers. The surface phase distribution map of the out-of-plane magnetic field ($H_z$) at the extrema of the first band –coinciding with the K point– is plotted in Fig. 1b. Opposite orbital-angular-momentum (OAM) states, which are phase singular points, around which the phase winds clockwise (cw) or counter-clockwise (ccw), are evenly distributed throughout the bulk, as indicated by the curled arrows in the inset. Fig. 1c shows the associated Berry curvature around the high symmetry K point of the Brillouin zone. The lack of Berry phase accumulation here resulting in zero Berry curvature, hence zero valley-Chern number, is a clear contrast to the case of valley PTIs. Detailed comparison between the studied triangular PhC and valley PhC is given in the supplementary materials.

Generally, one needs to break spatial inversion symmetry (SIS) to generate opposite, nonvanishing Berry curvature profiles at K/K' leading to valley Hall effect (VHE) and topological valley phase. [17, 18] The valley DOF and valley-contrast phenomena in electronic and Bosonic systems alike is linked to OAM states. [31, 32] This is in accordance with the fact the $C_{6v}$ lattice shown in Fig. 1 exhibits two OAM states that are evenly distributed throughout the bulk (i.e. cancel each other), whereas a topological valley phase, which is attainable by reduction to $C_{3v}$ symmetry, promotes one OAM state. Intuitively, the $C_{6v}$ symmetry doubly preserves the $C_{3v}$ rotational symmetry, [33] hence the PhC in Fig. 1a carries two OAM states at each valley (note that Fig. 1b shows phase map at K only). In the following, we show that it is possible to mimic some of the properties of topological valley structures in our $C_{6v}$ PhC, given we can distinguish the OAM states at the two valleys at specific locations in the lattice. Specifically, we propose virtually a location-dependent version of the VHE (i.e. LVHE) that does not rely on bulk Berry curvature.

**Valley selective coupling**

Fig. 1d indicates the locations in the bulk of the PhC (denoted by A and B), at which the OAM states occur. In the so-called LVHE, at each of these locations, the K valley exhibits the opposite characteristic (i.e. OAM state) to the K' valley. We consider a large but finite hexagonal region of the PhC (as in Fig. 1a) and place an external OAM source of a specific polarity at either A or B at the center of the PhC to probe its response. Fig. 1e shows the surface field map over the PhC when a cw (ccw) OAM source at A (B) excites the first band at K/K' valleys. We observe wave propagation towards only three corners of the hexagonal PhC region. These directions correspond to the K point in momentum space, which means that the cw (ccw) OAM polarity locally defines the K valley-polarized state at A (B). In contrast, Fig. 1f shows, when exciting ccw (cw) OAM source at A (B), waves propagate to the opposite directions of the previous case, indicating that ccw (cw) OAM polarity is locked to the K' valley at A (B). Meanwhile, the opposite response is observed for band 2 at the K and K' valleys (not shown), i.e. ccw/cw (cw/ccw) OAM states locally define the K/K' valley polarization at A (B), respectively. This proves that our PhC supports valley-contrast response that is analogous to conventional VHE in valley PTIs as far as only location A or B is concerned. Note that K and K' valleys are related by time-reversal symmetry (TRS); hence the fields at the K' valley could be deduced by applying a TRS operation to the fields at the K valley. TRS reverses the direction of the energy flux and the fields' phase rotation, and hence the polarity of the OAM states at any specific location in agreement with the above observations.

**Valley polarized edge states**

While the bulk-boundary correspondence principle prohibits topological edge states between a valley-projected topological phase [22, 28] and a topologically trivial phase, the domain wall between two crystals with half-integer valley-Chern numbers of opposite signs allows for topological kink states, i.e. VPEMs. This is different from the more general type of topological edge states that may occur at the exterior edge of a PTI. [28] Essentially, VPEMs are the result of the conservation of the binary valley DOF, [18, 34] hence VPEMs take place at the interface between structures with opposite OAMs. Where only a local region is concerned (e.g. A or B as discussed in the previous section), which is the case for a waveguide scenario, the LVHE upholds valley DOF. Therefore, despite the bandgap in the proposed PhC not arising from a broken SIS like in a valley PTI, we hypothesize it is possible to support VPEMs by enforcing opposite OAM polarities across a line defect in the PhC. We provide an analysis in the supplementary materials of how such states, which are trivial from the perspective of Berry curvature, result instead from an interplay between crystalline symmetries and finite boundary effects. A complete theoretical study of our structure can be found in [37], which shows how such states comprise a distinct topological phase, provided the gauge-dependent symmetries are maintained.

Fig. 2a plot the band diagram of the PhC with a line defect, showing guided edge modes that span the bulk bandgap between the first and second TE bulk bands. The line-defect waveguide (LDWG) is simply a one row of holes that is more densely packed than the rest of the PhC (see inset of Fig. 2e). Far from the defect region, the PhC on the two sides appear identical to each other. This is unlike valley PTIs where two different valley PhCs are interfaced to form a domain wall. Fig. 2b and c, which plot the $H_z$ phase and the Poynting vectors, respectively, across the LDWG, show a forward propagating wave concentrated along the defect interface with phase singular points appearing to the right and left sides that carry cw and ccw OAMs, respectively. This corresponds to the mode denoted by a dark purple color in Fig. 2a at mid-gap. Like valley PTIs, the translation symmetry of the lattice here is preserved along the direction parallel to LDWG (denoted by $k_{//}$ in Fig. 2d). That is, the line defect is aligned with the ΓK and ΓK′ directions of the PhC. The linkage between the propagation direction (ΓK or ΓK′ direction) and the specific OAM states across the line defect gives rise to direction-locked propagation, meaning the edge modes are valley polarized (i.e. VPEMs). Since the band diagram is symmetric with respect to the wavevector, there exists a pair of counter-propagating edge states with opposite valley-polarizations; the mode denoted by light magenta color in Fig. 2a is associated with a backward propagating wave, which has the reversed orientation of OAM states.

Fig. 2e shows an LDWG with multiple sharp turns excited by an OAM source (marked by yellow star) at frequency at midgap of the PhC. The simulated surface field map shows that the edge mode is transmitted unidirectionally to the right of the sample and propagates through every bent segment without reflection. As is the case of a valley PTI, a zigzag-shaped pathway along the ΓK inclination, which conserves the valley DOF, can evidently support VPEMs in our PhC. We further prove the nature of these edge modes and the reason for their robustness against sharp bends by testing the wave routing through a magic-T junction, as shown in Fig. 2f. The surface field map shows that when a wave is injected from port 2, it is routed into ports 1 and 3 but, counterintuitively, not port 4. This can be explained by the edge mode being valley polarized. As marked in Fig. 3d, the guided mode in the input port 1 belongs to the K valley, which is of the same valley polarization as that of the output ports 1 and 3. On the other hand, the valley polarization of the output port 4 belongs to the K′ valley, so light cannot be coupled into this port. As such, the edge states in our PhC indeed share the same origin as the VHE and inherit similar features to VPEMs in valley PTIs.

**Optical Measurements**

We fabricated LVHE-based LDWG devices on a standard SOI wafer with straight, zigzag, and double zigzag pathways (see Methods) as shown by the scanning-electron-microscope (SEM) images in Fig. 3a. We chose a lattice constant of 380 nm, air hole diameter of 160 nm and slab thickness of 220 nm. The separation distance between adjacent holes at the line defect was 60 nm and the devices were covered with 3μm buried oxide layer. This gives a TE bandgap spanning the wavelength range of 1514nm to 1595nm. The PhC bandgap is in the guide part of the band diagram (i.e. below the light line), hence the VPEMs will be confined in the plane of the PhC slab.

Fig. 3b shows the measured transmission spectra of the proposed waveguides in the wavelength range of 1530–1565nm, which is limited by the grating coupler performance used for testing (see Methods). The measurement results show high transmittance that is comparable for the three interfaces, as expected of VPEMs (note the values are in linear scale and include loss by the grating coupler, hence the bell-shaped response). We attribute the recorded insertion losses at corners to scattering into plane waves in the high-dielectric $SiO_2$ substrate and buried cladding layer (see Methods). For reference, similar waveguide devices using a conventional valley PTI [24] were fabricated and tested, as shown in Fig. 3c. The measured transmission spectra are comparable to the results from the LVHE-based LDWGs. In addition, we fabricated and tested typical LDWGs [35] with similar sharp bends, as shown in Fig. 2d. As expected, light transmission is greatly deteriorated due to the sharp bends in contrast to the previous two types of devices.

**Conclusion**

We have presented a new paradigm for realizing valley kink states and experimentally confirmed their robust light transmission through sharp bends using an SOI slab at telecommunication wavelengths. These modes share similar characteristics to topological modes in VHE-based PTIs albeit happening in a bandgap PhC with zero valley-Chern number. Instead, as pointed out recently in [37], a distinct topological phase characterizes our structure based on symmetry indicators or alternatively by computing charge polarization, where displacement of the Wannier center from the center of the (real space) unit cell indicates a charge imbalance that is compensated by edge states on a finite sample. More simply, the VPEMs here can be understood as the product of a line defect in a lattice with LVHE (a general feature of $C_{6v}$ point symmetry), where the defect causes a phase discontinuity in spatially-varying OAM states distribution. Our results reveal the role of interface effects in forming gapless valley kink states and present a new mean to create valley DOF-based-like devices. This includes new opportunities to develop low-loss compact delay lines [27], slow-light optical buffers, and lasers [26, 30]. Lastly, the waveguiding phenomenon demonstrated here is applicable not only to the electromagnetic spectrum and various optical systems, but also to other classical wave systems such as plasmonics [38], mechanics and acoustics [39].

**Methods**

Fabrication:

The PhCs were fabricated on an SOI wafer, with a 220-nm-thick silicon (Si) device layer over silica ($SiO_2$). The device patterns were defined by high-resolution e-beam lithography and then transferred to the silicon device layer by plasma dry etching. Subsequently, a 3-µm-thick buried $SiO_2$ cladding layer was deposited on top of the Si layer for protection using PECVD process. We fabricated the reference valley PTI on the same SOI wafer using the same process. The associated PhC has a honeycomb lattice of circular air holes with diameters of 160 nm and 80 nm for A and B sites, respectively, and a periodicity of 380 nm. The layout of the waveguide interface was chosen to be the same as reported in Ref. 24. In addition, we fabricated a conventional LDWG for comparison on the same SOI wafer using the same parameters as our LVHE PhC. This LDWG's width was 658 nm, which is defined as the distance between the centers of the air holes nearest the waveguide.

Measurement:

To measure the transmission spectra of the fabricated devices, light from a tunable semiconductor laser was coupled via a single-mode fiber to the SOI waveguides through an integrated grating coupler, which had a bandwidth of $\approx$ 35 nm. The TE-polarized continuous waves at the telecommunication wavelength were coupled to a 1.55µm-width input rectangular waveguide and then launched into the PhC sample. We used a linear taper to connect the waveguide of 500 nm width at the grating coupler to the waveguide of 1550 nm width at the PhCs' facets. After passing through the PhC, the propagating wave was coupled to the output rectangular waveguide and then collected by another grating coupler. The transmission spectra of the devices were obtained by sweeping the laser wavelength and simultaneously measuring the transmitted signal using a high-sensitivity optical power meter. Note that the dimensions of the strip waveguides that couple into and out of the PhC devices was chosen the same as given in Ref. 24 without further optimization, which explains the enhanced coupling efficiency (higher transmittance) of the corresponding devices compared to the proposed LVHE-based LDWGs.

Numerical simulation:

All the numerical simulation results, including band diagrams, eigen-field patterns and optical transmission data in this work were retrieved from full-wave electromagnetic simulation using the commercial finite element method solver software Ansys HFSS. Floquet periodic boundaries were assigned for unit cell and supercell band diagram calculations, while radiation boundary was assigned for the outer boundaries of the structure in case of the frequency domain driven-mode studies. To generate an OAM excitation source, we used a circular arrangement of a group of $H_z$ dipole-point sources with same magnitudes and incrementally increasing phases as illustrated in Fig. 1d. For the Berry curvature calculation [36], the complex out-of-plane H-field with magnitude and phase information in the real-space domain is exported from HFSS simulations for each wavevector for the integration.

**Acknowledgements**

This work was supported in part by DARPA grant W911NF-17-1-0580.

**Conflict of Interest**

The authors declare no conflict of interest.

**Contributions**

D.J.B conceived the idea, performed the numerical simulations and experiments, and analyzed the data. D.F.S supervised the project and interpreted the results. D.J.B prepared the manuscript and D.F.S provided feedback.

**Data Availability**

The data that support the findings of this study are available from the corresponding author upon reasonable request.

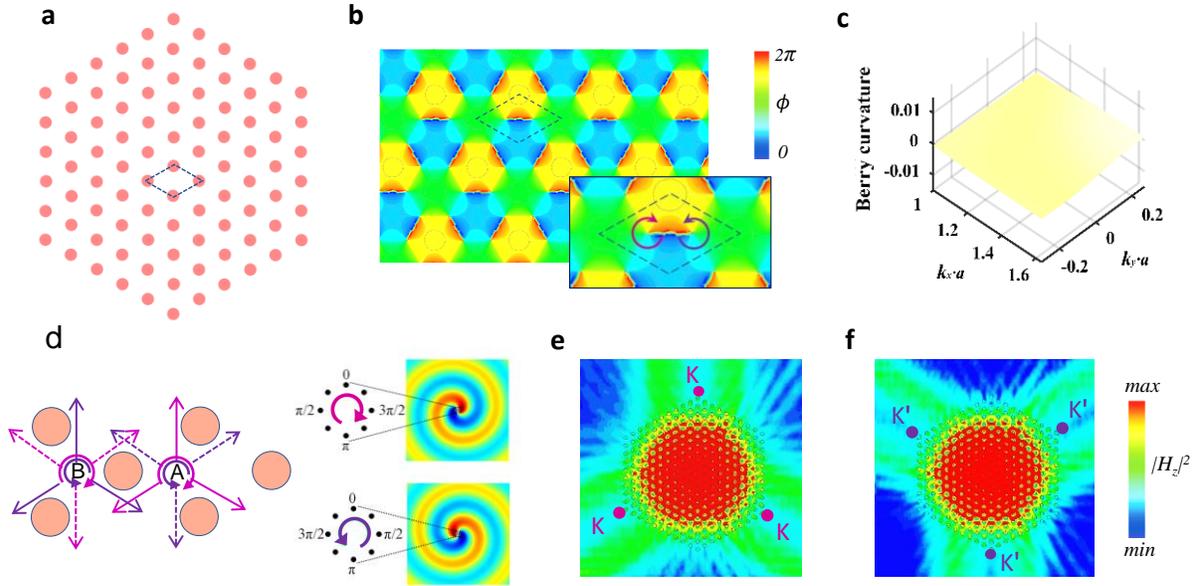

**Fig. 1: Spatial distribution of OAM states and valley-contrast response based on LVHE in a triangular lattice PhC. a,** Two-dimensional triangular lattice of circular air holes in a dielectric. Due to the $C_{6v}$ symmetry, the lattice does not qualify as a valley PhC. **b,** Simulated surface phase map of $H_z$ field at the K valley of the fundamental TE band of the PhC. The map shows phase vortices, which correspond to OAM states of opposite polarities (as shown in the inset) that appear equally through the bulk. **c,** Wilson loop diagram of the PhC showing no signature of any topological order. **d**, Points A and B specify the locations, at which a source carrying an OAM state would excite a wave with maximum directionality. **e**, Surface map of $H$-field intensity over the PhC region illustrated in **a**, showing wave propagation towards K valley direction when excited by cw OAM state at point A or ccw OAM state at point B. **f**, same map as in **e** but showing wave propagation towards K′ valley direction when excited by ccw OAM state at point A or cw OAM state at point B.

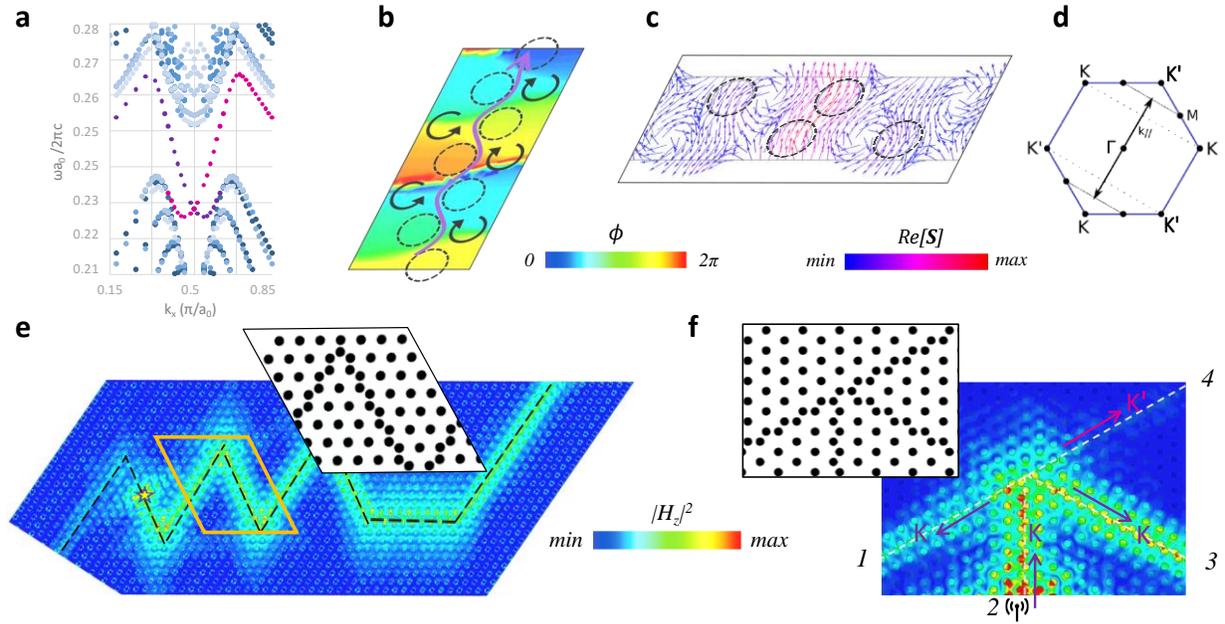

**Fig. 2: Schematics and characteristics of a line defect in a triangular PhC supporting VPEMs based on LVHE. a,** Band diagram of TE modes of the 2D air holes triangular PhC containing a line defect (illustrated in **b** and **e**), showing the dispersion of guided edge modes within the bulk bandgap. The structure parameters are as follows: air holes diameter of 160 nm, periodicity of 380 nm, holes' separation at the interface of 60 nm, and silicon dielectric, b, Surface phase map of Hz filed across the line defect, which is formed by closer packing of holes along one row of holes compared to the rest of the PhC. The map shows opposite OAM states (denoted by curled arrows) across the defect that are associated with upward traveling wave, i.e. a VPEM. **c**, Plot of the Poynting (*S*) vectors corresponding to the results in **b**, showing energy flow along the line defect and flux vortices on the two sides in accordance with the OAM states in **b**. **d**, Illustration of the Brillouin zone of the PhC indicating the orientation k// parallel to the line defect, showing that the translation symmetry of the PhC in ΓK and ΓK' directions is preserved at along the defect. **e,** Surface map of *H*-field intensity over the PhC showing unidirectional excitation of the VPEM when using a proper polarized source (denoted by yellow star), and reflection-free transmission along a meandering interface pathway (marked by dashed line). **f,** Planar magic-T junction showing VPEMs, routing each polarization state towards its corresponding valley direction analogous to spin-momentum locking.

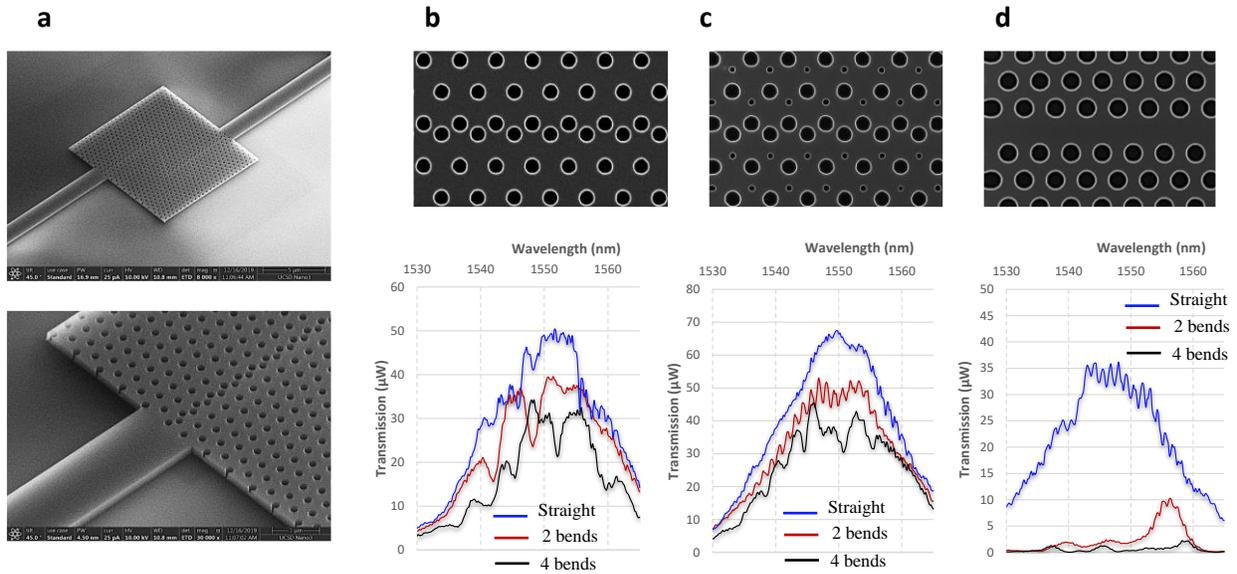

**Fig. 3: Experimental observation of robust VPEMs in an SOI triangular PhC slab. a,** Perspective views of scanning-electron-microscope (SEM) images of the fabricated PhC waveguide device (LVHE-based LDWG) with the zigzag pathway on SOI slab. The PhC is coupled directly by a standard rectangular strip silicon waveguide. **b**, Measurements of transmission spectra of the proposed LDWG in case of straight, zigzag (two 120º bends) and double zigzag (four 120º bends) waveguide pathways. **c**, **d** Same as **b** but for VHE-based PTI and common LDWG. Only in case of the LVHE and VHE devices, the spectra in the bandgap maintain high transmittance for a sharp-bending geometry.